\documentclass[a4paper,english]{article}
\usepackage[T1]{fontenc}
\usepackage[latin9]{inputenc}
\usepackage{float}
\usepackage{textcomp}
\usepackage{amsthm}
\usepackage{amsmath}
\usepackage{graphicx}
\usepackage{amssymb}

\makeatletter

\pdfpageheight\paperheight
\pdfpagewidth\paperwidth

\providecommand{\tabularnewline}{\\}

\numberwithin{equation}{section}
\numberwithin{figure}{section}

\makeatother

\usepackage{babel}

\begin{document}

\title{\noindent Control of the repetitive firing in the squid giant axon
using electrical fields }

\author{\noindent R. Ozgur Doruk }

\maketitle
\noindent Middle East Technical University \textendash{} North Cyprus
Campus, Kalkanl\i{}, Güzelyurt, Mersin 10, TURKEY (TR) e \textendash{}
mail: rdoruk@metu.edu.tr
\begin{abstract}
\noindent In this research, the aim is to develop a repetitive firing
stopper mechanism using electrical fields exerted on the fiber. The
Hodgkin \textendash{} Huxley nerve fiber model is used for modeling
the membrane potential behavior. The repetitive firing of the nerve
fiber can be stopped using approaches based on the control theory
where the nonlinear Hodgkin \textendash{} Huxley model is used to
achieve this goal. The effects of the electrical field are considered
as an additive quantity over the equilibrium potentials of the cell
membrane channels 

\end{abstract}

\section{\noindent Introduction }

\noindent The repetitive firing of the nerve fibers in the living
beings is an important subject of the biophysical research since the
phenomenon has various meanings. It may have a role in several diseases
like epilepsy {[}Pellock et.al (2001){]} but have also a role in the
formation of heart beats where the oscillation frequency is in fact
represents the heart rate. The Hodgkin \textendash{} Huxley nerve
fiber model {[}Hodgkin et.al (1952){]} provides a very useful framework
for simulation based neurophysiologic studies where the variation
of the membrane potential is modeled basically as a fourth order nonlinear
differential equation. The basic model in {[}Hodgkin et.al (1952){]}
allows the external current injection as the system input whereas
an additional cell membrane potential can also be considered as an
input which is the case in {[}Wang et.al. (2004){]}. The electrical
fields around the nerve cell membranes can induce small electric potentials
which may lead to different behaviors as stated in {[}Wang et.al.
(2004), Kotnik et.al (1998){]} however the potential induction mechanism
can also be used as a repetitive firing control mechanism which is
the main point of this research. The repetitive firing condition on
the Hodgkin \textendash{} Huxley nerve fiber model is detected using
a bifurcation discovery tool like MATCONT {[}Dhooge et.al. (2003){]}
and the washout filter {[}Hassouneh et.al. (2004), Wang et.al. (2000){]}
and projective control {[}Medanic (1978){]} is integrated to derive
a membrane additive voltage profile. This voltage profile is converted
into an electrical field requirement profile using an approach derived
from {[}Kotnik et.al. (1998){]}. Using electrical energy in treatment
of neurological disorders is a wide part of neurophysiology research
where one of the most prominent applications is the deep brain stimulation
which is used in the treatment of Parkinson\textquoteright{}s disease
{[}Moro et.al. (2006){]} and depression {[}Mayberg et.al. (2005){]}.
The mechanism is that a device implanted in the brain tissue sending
electrical pulses to the brain. This can be taught as a current injection
system whereas in this research the mechanism of electric field induction
is used to obtain a response from the neurological tissue. The research
can be a base point for further advancements in electrical field stimulation
of the central nervous system. Using electrical energy in treatment
of neurological disorders is a wide part of neurophysiology research
where one of the most prominent applications is the deep brain stimulation
which is used in the treatment of Parkinson\textquoteright{}s disease
{[}Moro et.al. (2006){]} and depression {[}Mayberg et.al. (2005){]}.
The mechanism is that a device implanted in the brain tissue sending
electrical pulses to the brain. This can be taught as a current injection
system whereas in this research the mechanism of electric field induction
is used to obtain a response from the neurological tissue. The research
can be a base point for further advancements in electrical field stimulation
of the central nervous system. In this paper, first of all the Hodgkin
\textendash{} Huxley model of nerve fibers together with the electrical
field effects are reviewed and then the methods of projective control
theory and washout filters are presented. Next the approach of this
research is described in detail.

\section{\noindent Materials and Methods }

\subsection{\noindent Hodgkin Huxley Model and Electrical Field Interaction }

\noindent Hodgkin \textendash{} Huxley model nerve fiber dynamics
{[}Hodgkin et.al. (1952){]} is a fourth order differential equation
derived for modeling the behavior of the membrane potential as shown
below: 

\noindent \begin{eqnarray}
C_{m}\frac{dV}{dt} & = & -g_{Na}m^{3}h\left(V-V_{Na}\right)-g_{K}n^{4}\left(V-V_{K}\right)-g_{L}\left(V-V_{L}\right)+I_{ext}\label{hodgkin_huxley}\end{eqnarray}

\noindent \[
\frac{dn}{dt}=\alpha_{n}\left(1-n\right)-\beta_{n}n\]

\noindent \[
\frac{dm}{dt}=\alpha_{m}\left(1-m\right)-\beta_{m}m\]

\noindent \[
\frac{dh}{dt}=\alpha_{h}\left(1-h\right)-\beta_{h}h\]

\noindent \[
\alpha_{n}=\frac{0.1-0.01V}{exp\left(1-0.1V\right)-1},\,\beta_{n}=0.125exp\left(-\frac{V}{80}\right)\]

\noindent \[
\alpha_{m}=\frac{2.5-0.1V}{exp\left(2.5-0.1V\right)-1},\,\beta_{m}=4exp\left(-\frac{V}{18}\right)\]

\noindent \[
\alpha_{h}=0.07exp\left(-\frac{V}{20}\right),\,\beta_{h}=\frac{1}{exp\left(3-0.1V\right)-1}\]

\noindent The definitions of the parameters are: 

\noindent V: the displacement of the cell membrane potential from
its resting value in mV 

\noindent n : is a dimensional variable which can vary between 0 and
1 representing the proportion of activating molecules of the potassium
channel 

\noindent m : same type of variable as n which represents the proportion
of the activating molecules of the sodium channel 

\noindent h : same type of variable as n which represents the proportion
of the inactivating molecules of the sodium channel 

\noindent $I_{ext}$ : external current injection in $\frac{\mu A}{cm^{2}}$

\noindent $g_{Na}$ : sodium channel conductance in $\frac{mS}{cm^{2}}$

\noindent $g_{K}$ : potassium channel conductance in $\frac{mS}{cm^{2}}$

\noindent $g_{L}$: leakage conductance in $\frac{mS}{cm^{2}}$ due
to the chloride and other ions which leads to a leaking current flow.
Because of that the behaviour is modeled as a conductance. 

\noindent $C_{m}$: membrane capacitance in $\frac{\mu F}{cm^{2}}$

\noindent $V_{Na}:$ sodium channel resting potential in mV 

\noindent $V_{K}:$ potassium channel resting potential in mV

\noindent $V_{L}:$ the level of potential where the leakage current
reduces to zero 

\noindent In the experimentation performed by Alan Hodgkin and his
colleagues the nominal values determined are:

\noindent \begin{equation}
g_{Na}=36\frac{mS}{cm^{2}}\label{HH_parameters}\end{equation}

\noindent \[
g_{K}=120\frac{mS}{cm^{2}}\]

\noindent \[
g_{L}=0.3\frac{mS}{cm^{2}}\]

\noindent \[
V_{K}=-12mV\]

\noindent \[
V_{Na}=115mV\]

\noindent \[
V_{L}=10.613mV\]

\noindent \[
C_{m}=0.91\frac{\mu F}{cm^{2}}\]

\noindent In this work the external current injection is set to zero
by default. It is to be used as an external stimulus for stabilizing
the neuron with the washout filter. The external effect input to the
model in our case is the external voltage induction on the cell membrane
by the electric fields which is added to (2.1) as: 

\noindent \begin{eqnarray}
C_{m}\frac{dV}{dt} & = & -g_{Na}m^{3}h\left(V+V_{E}-V_{Na}\right)-g_{K}n^{4}\left(V+V_{E}-V_{K}\right)-g_{L}\left(V+V_{E}-V_{L}\right)+I_{ext}\label{HH_Elec_Fields}\end{eqnarray}

\noindent \[
\frac{dn}{dt}=\alpha_{n}\left(1-n\right)-\beta_{n}n\]

\noindent \[
\frac{dm}{dt}=\alpha_{m}\left(1-m\right)-\beta_{m}m\]

\noindent \[
\frac{dh}{dt}=\alpha_{h}\left(1-h\right)-\beta_{h}h\]

\noindent \[
\alpha_{n}=\frac{0.1-0.01V}{exp\left(1-0.1V\right)-1},\,\beta_{n}=0.125exp\left(-\frac{V}{80}\right)\]

\noindent \[
\alpha_{m}=\frac{2.5-0.1V}{exp\left(2.5-0.1V\right)-1},\,\beta_{m}=4exp\left(-\frac{V}{18}\right)\]

\noindent \[
\alpha_{h}=0.07exp\left(-\frac{V}{20}\right),\,\beta_{h}=\frac{1}{exp\left(3-0.1V\right)-1}\]

The term $V_{E}$ is the potential induced on the cell membrane by
any external effect which is the electrical field in this case. If
this is due to a time varying electrical field relationship between
$V_{E}$ and the exerting electrical field E is obtained from the
solution of Laplace\textquoteright{}s partial differential equation
which is used by {[}Kotnik, 1998{]} to obtain the following:

\begin{equation}
V_{E}=F\left(s\right)E\left(s\right)Rcos\left(\theta\right)\label{HH_E(s)}\end{equation}

where $E\left(s\right)$ is the electrical field time course defined
in the Laplace domain, R is the radius of the cell soma and $\theta$
is the directional angle of the electrical field exerted upon the
nerve fibers. Finally $F\left(s\right)$ is the transfer function
defined as:

\begin{equation}
F\left(s\right)=\frac{a_{1}s^{2}+a_{2}s+a_{3}}{b_{1}s^{2}+b_{2}s+b_{3}}\label{HH_F(S)}\end{equation}

with its coefficients:

\inputencoding{latin1}\begin{equation}\begin{array}{l}  a_1  = 3d\lambda _o \left( {\lambda _i \left( {3R^2  - 3dR + d^2 } \right) + \lambda _m \left( {3dR - d^2 } \right)} \right) \\   a_2  = 3d\left( {\left( {\lambda _i \varepsilon _o  + \lambda _o \varepsilon _i } \right)\left( {3R^2  - 3dR + d^2 } \right) + \left( {\lambda _m \varepsilon _o  + \lambda _o \varepsilon _m } \right)\left( {3dR - d^2 } \right)} \right) \\   a_3  = 3d\varepsilon _o \left( {\varepsilon _i \left( {3R^2  - 3dR + d^2 } \right) + \varepsilon _m \left( {3dR - d^2 } \right)} \right) \\   b_1  = 2R^3 \left( {\lambda _m  + 2\lambda _o } \right)\left( {\lambda _m  + \frac{1}{2}\lambda _i } \right) + 2\left( {R - d} \right)^3 \left( {\lambda _m  - \lambda _o } \right)\left( {\lambda _i  - \lambda _m } \right) \\   b_2  = 2R^3 \left( {\lambda _i \left( {\frac{1}{2}\varepsilon _m  + \varepsilon _o } \right) + \lambda _m \left( {\frac{1}{2}\varepsilon _i  + 2\varepsilon _m  + 2\varepsilon _o } \right) + \lambda _o \left( {\varepsilon _i  + 2\varepsilon _m } \right)} \right) + 2\left( {R - d} \right)^3  \\    \times \left( {\lambda _i \left( {\varepsilon _m  - \varepsilon _o } \right) + \lambda _m \left( {\varepsilon _i  - 2\varepsilon _m  + \varepsilon _o } \right) - \lambda _o \left( {\varepsilon _i  - \varepsilon _m } \right)} \right) \\   b_3  = 2R^3 \left( {\varepsilon _m  + 2\varepsilon _o } \right)\left( {\varepsilon _m  + \frac{1}{2}\varepsilon _i } \right) + 2\left( {R - d} \right)^3 \left( {\varepsilon _m  - \varepsilon _o } \right)\left( {\varepsilon _i  - \varepsilon _m } \right) \\   \end{array} \end{equation}

The parameters of (2.6) are defined in the following table in reference
to {[}Kotnik, 1998{]}:

\begin{table}[H]
\caption{\selectlanguage{english}%
The values of electrical parameters for the considered cell membrane
(values are from Kotnik et.al)}

\noindent \centering{}\begin{tabular}{|c|c|c|}
\hline 
\selectlanguage{english}%
\inputencoding{latin9}Parameter  & \selectlanguage{english}%
Definition  & \selectlanguage{english}%
\inputencoding{latin9}Value \tabularnewline
\hline
\hline 
\selectlanguage{english}%
$\lambda _i $  & \selectlanguage{english}%
\inputencoding{latin9}Cytoplasmic conductivity  & \selectlanguage{english}%
$0.3\,S \cdot m^{ - 1} $ \tabularnewline
\hline 
\selectlanguage{english}%
\inputencoding{latin9}$\varepsilon _i $  & \selectlanguage{english}%
Cytoplasmic permittivity  & \selectlanguage{english}%
\inputencoding{latin9}$7.1 \cdot 10^{ - 10} \,A \cdot s \cdot V^{ - 1}  \cdot m^{ - 1} $ \tabularnewline
\hline 
\selectlanguage{english}%
$\lambda _m $  & \selectlanguage{english}%
\inputencoding{latin9}Membrane conductivity  & \selectlanguage{english}%
$3 \cdot 10^{ - 7} \,S \cdot m^{ - 1} $ \tabularnewline
\hline 
\selectlanguage{english}%
\inputencoding{latin9}$\varepsilon _m $  & \selectlanguage{english}%
Membrane permittivity  & \selectlanguage{english}%
\inputencoding{latin9}$4.4 \cdot 10^{ - 11} \,A \cdot s \cdot V^{ - 1}  \cdot m^{ - 1} $ \tabularnewline
\hline 
\selectlanguage{english}%
$\lambda _o $  & \selectlanguage{english}%
\inputencoding{latin9}Extracellular medium conductivity  & \selectlanguage{english}%
$3 \cdot 10^{ - 1} \,S \cdot m^{ - 1} $ \tabularnewline
\hline 
\selectlanguage{english}%
\inputencoding{latin9}$\varepsilon _o $  & \selectlanguage{english}%
Extracellular medium permittivity  & \selectlanguage{english}%
\inputencoding{latin9}$7.1 \cdot 10^{ - 10} \,A \cdot s \cdot V^{ - 1}  \cdot m^{ - 1} $ \tabularnewline
\hline 
\selectlanguage{english}%
$R$  & \selectlanguage{english}%
\inputencoding{latin9}Cell soma radius  & \selectlanguage{english}%
$10\,\mu m$ \tabularnewline
\hline 
\selectlanguage{english}%
\inputencoding{latin9}$d$  & \selectlanguage{english}%
Membrane thickness  & \selectlanguage{english}%
\inputencoding{latin9}$5\,nm$ \tabularnewline
\hline
\end{tabular}
\end{table}

\selectlanguage{english}%
\inputencoding{latin9}The relation in (2.4) is derived using admittivity
operators. The details of derivations are given in the relevant research
{[}Kotnik et.al. (1998){]}. In order to obtain the level of electric
field required for the intended operation the following can be derived: 

\begin{equation}E\left( s \right) = \frac{1}{{R\cos \left( \theta  \right)}}\Phi \left( s \right)V_E  \end{equation} 

where,

\begin{equation} \Phi \left( s \right) = \frac{{b_1 s^2  + b_2 s + b_3 }}{{a_1 s^2  + a_2 s + a_3 }} \end{equation} 

The above can be written since the order of the numerator and denominator
in (2.5) is same where the inversion does not change the properness
of the function. For maximum efficiency the directional angle of the
field $\theta$ should be zero.

\subsection{Projective Control Theory}

The projective control approach is a linear control methodology that
derives an output feedback controller from a full state feedback design.
The transformation from the full state feedback to the output feedback
is performed through the orthogonal projection which operates on the
closed loop eigenspectrum of the full state feedback design. Before
going into the detail of the projective control approach a review
of full state feedback control approach is given for convenience.
Consider a linear system depicted by:

\begin{equation}\label{linear_system} {\bf{\dot x}} = {\bf{Ax}} + {\bf{Bu}} \end{equation} 

where\inputencoding{latin1}{ \[ {\mathbf{x}} \in \mathbb{R}^n ,\,{\mathbf{u}} \in \mathbb{R}^m ,\,{\mathbf{A}} \in \mathbb{R}^{n \times n} ,\,{\mathbf{B}} \in \mathbb{R}^{n \times m}  \] }

If one designs a full state feedback controller as \[ {\mathbf{u}} =  - {\mathbf{Kx}} \] 
the closed loop dynamics is now \[ {\mathbf{\dot x}} = \left( {{\mathbf{A}} - {\mathbf{BK}}} \right){\mathbf{x}} \] 
The eigenspectrum of this closed loop dynamics is defined by the equation
shown below:

\begin{equation} \left( {{\mathbf{A}} - {\mathbf{BK}}} \right){\mathbf{V}} = {\mathbf{V\Lambda }} \end{equation} 

where$\Lambda$ is the diagonal matrix with the entries of eigenvalues
of the closed loop dynamics ( or the eigenvalues of ${\mathbf{A}} - {\mathbf{BK}} $ 
) and is the matrix of the corresponding eigenvectors. If one is thinking
of making a static output feedback from ${\mathbf{y}} = {\mathbf{Cx}}$ 
$\left( {{\mathbf{C}} \in \mathbb{R}^{q \times n} ,\,{\mathbf{y}} \in \mathbb{R}^q } \right)$
as $ {\mathbf{u}} =  - {\mathbf{K}}_o {\mathbf{y}} =  - {\mathbf{K}}_o {\mathbf{Cx}} $ .
If this feedback is assumed to retain $q$ eigenvalues out of $\Lambda$
(denote this as $\Lambda_{q}$ and corresponding eigenvectors from
$V$ as $V_{q}$ ) the following should also be written:

\begin{equation} \left( {{\mathbf{A}} - {\mathbf{BKC}}} \right){\mathbf{V}}_q  = {\mathbf{V}}_q {\mathbf{\Lambda }}_q  \end{equation} 

Combining (2.10) and (2.11) the following can be written:

\[ \left( {{\mathbf{A}} - {\mathbf{BK}}} \right){\mathbf{V}} = \left( {{\mathbf{A}} - {\mathbf{BK}}_o {\mathbf{C}}} \right){\mathbf{V}}_q  \] 

This allows one to solve for \[ {\mathbf{K}}_o  \]  as:

\begin{equation}\mathbf{K}_o  = {\mathbf{KV}}_q \left( {{\mathbf{CV}}_q } \right)^{ - 1} \end{equation} 

For the nonlinear systems the model should be linearized using the
Jacobian (or the Taylor series).

\subsection{Linearization and Bifurcation}

As it is just stated the projective control approach requires linearization
of the nonlinear model. For the case of a standard nonlinear system
representation like ${\mathbf{\dot x}} = {\mathbf{f}}\left( {{\mathbf{x}},\,{\mathbf{u}}} \right)$ 
the following equations are written:

\begin{equation}\begin{gathered}   {\mathbf{\dot x}} = {\mathbf{Ax}} + {\mathbf{Bu}} + {\mathbf{H}}\left( {{\mathbf{x}},\,{\mathbf{u}}} \right) \hfill \\   {\mathbf{A}} = \left. {\frac{{\partial {\mathbf{f}}}} {{\partial {\mathbf{x}}}}} \right|_{{\mathbf{x}} = {\mathbf{x}}_0 }  \hfill \\   {\mathbf{B}} = \left. {\frac{{\partial {\mathbf{f}}}} {{\partial {\mathbf{u}}}}} \right|_{{\mathbf{u}} = {\mathbf{u}}_0 }  \hfill \\  \end{gathered}  \end{equation}

where ${\mathbf{x}} \in \mathbb{R}^n $  is the state of the system,
${\mathbf{u}} \in \mathbb{R}^m $  is the input and $ {\mathbf{H}}\left( {{\mathbf{x}},\,{\mathbf{u}}} \right) $ 
is the higher order nonlinear terms and ${\mathbf{x}}_0 $  is the
equilibrium point. The eigenvalues of the matrix characterizes the
stability of ${\mathbf{x}}_0 $ . If the instability of the system
(2.13) is due to a single pair of complex conjugate eigenvalues the
resultant event is called as Hopf bifurcation. On the other hand if
there is a pair of eigenvalues with the same values and opposite signs
the resultant issue is called as the saddle node bifurcation. For
the Hodgkin \textendash{} Huxley model the repetitive firing in non
\textendash{} stimulated (current or other external effects) is a
result of the bifurcation phenomenon. The bifurcation characteristics
of the system can be changed by using specially designed controllers.
However, many of the controllers lead to a change in the position
of the equilibrium point. This is not a problem in general but if
the equilibrium point should not be changed one has to incorporate
a washout filter which blocks the steady state (DC) inputs to the
system. The washout filter itself is a linear high \textendash{} pass
filter which can be mathematically defined as shown below:

\begin{equation} \begin{gathered}   {\mathbf{\dot z}} = {\mathbf{A}}_{\mathbf{w}} {\mathbf{z + B}}_{\mathbf{w}} {\mathbf{\xi }} \hfill \\   {\mathbf{\psi  = A}}_{\mathbf{w}} {\mathbf{z + B}}_{\mathbf{w}} {\mathbf{\xi }} \hfill \\  \end{gathered}  \end{equation}

where ${\mathbf{z}} \in \mathbb{R}^p $ , ${\mathbf{\psi }} \in \mathbb{R}^p $, ${\mathbf{\xi }} \in \mathbb{R}^p $ , ${\mathbf{A}}_w  \in \mathbb{R}^{p \times p} $ and ${\mathbf{B}}_w  \in \mathbb{R}^{p \times p} $ 

If the above dynamics is expressed in the Laplace domain the following
result can be obtained:

\begin{equation} {\mathbf{G}}\left( s \right) = \frac{{{\mathbf{\psi }}\left( s \right)}} {{{\mathbf{\xi }}\left( s \right)}} = {\mathbf{A}}_{\mathbf{w}} \left[ {s{\mathbf{I}}_{p \times p}  - {\mathbf{A}}_{\mathbf{w}} } \right]^{ - 1} {\mathbf{B}}_{\mathbf{w}}  + {\mathbf{B}}_{\mathbf{w}}  \end{equation}

where $ {\mathbf{G}}\left( s \right) $  is the multi \textendash{}
input and multi \textendash{} output transfer function from the input
${\mathbf{\xi }}$  to output $ {\mathbf{\psi }} $ . It is clearly
noted that the following limit tends to zero. 

\begin{equation}\mathop {\lim }\limits_{s \to 0} {\mathbf{G}}\left( s \right) \to 0 \end{equation} 

The above result means that the system in (2.14) blocks the steady
state inputs. This property is the core part of the control mechanisms
aided with a washout filter.

\subsection{Hodgkin \textendash{} Huxley Model and application}

For the Hodgkin \textendash{} Huxley model in consideration first
of all a linearization should be done. So the linearization will lead
to the linearized model in (\ref{linear_system}) with the matrix$A$
as:

\begin{equation} {\mathbf{A}} = \left. {\left[ {\begin{array}{*{20}c}    {\frac{{\partial \dot V}} {{\partial V}}} & {\frac{{\partial \dot V}} {{\partial n}}} & {\frac{{\partial \dot V}} {{\partial m}}} & {\frac{{\partial \dot V}} {{\partial h}}}  \\    {\frac{{\partial \dot n}} {{\partial V}}} & {\frac{{\partial \dot n}} {{\partial n}}} & 0 & 0  \\    {\frac{{\partial \dot m}} {{\partial V}}} & 0 & {\frac{{\partial \dot m}} {{\partial m}}} & 0  \\    {\frac{{\partial \dot h}} {{\partial V}}} & 0 & 0 & {\frac{{\partial \dot h}} {{\partial h}}}  \\
 \end{array} } \right]} \right|_{V_E  = 0} ,\,{\mathbf{x}} = \left[ {\begin{array}{*{20}c}    V  \\    n  \\    m  \\    h  \\
 \end{array} } \right] \end{equation}

\begin{equation}\begin{gathered}   \frac{{\partial \dot V}} {{\partial V}} = \frac{1} {{C_m }}\left( { - g_{Na} m^3 h - g_K n^4  - g_L } \right) \hfill \\   \frac{{\partial \dot V}} {{\partial n}} =  - \frac{4} {{C_m }}g_K n^3 \left( {V - V_K } \right) \hfill \\   \frac{{\partial \dot V}} {{\partial m}} =  - \frac{3} {{C_m }}g_{Na} m^2 h\left( {V - V_{Na} } \right) \hfill \\   \frac{{\partial \dot V}} {{\partial h}} =  - \frac{1} {{C_m }}g_{Na} m^3 \left( {V - V_{Na} } \right) \hfill \\  \end{gathered}  \end{equation}

\begin{equation}\begin{gathered}   \frac{{\partial \dot n}} {{\partial V}} =  - 1/100/(\exp (1 - 1/10V) - 1)(1 - n) +  \hfill \\   1/10 \times (1/10 - 1/100V)/(\exp (1 - 1/10V) - 1)^2  \times (1 - n)\exp (1 - 1/10V) + 1/640 \times \exp ( - 1/80V)n \hfill \\   \frac{{\partial \dot m}} {{\partial V}} =  - 1/10/(\exp (5/2 - 1/10V) - 1)(1 - m) +  \hfill \\   1/10(5/2 - 1/10V)/(\exp (5/2 - 1/10V) - 1)^2 (1 - m) \times \exp (5/2 - 1/10V) + 2/9\exp ( - 1/18V)m \hfill \\   \frac{{\partial \dot h}} {{\partial V}} =  - 7/2000\exp ( - 1/20V)(1 - h) - 1/10/(\exp (3 - 1/10V) + 1)^2 h \times \exp (3 - 1/10V) \hfill \\   \frac{{\partial \dot n}} {{\partial n}} =  - \left( {\alpha _n  + \beta _n } \right) \hfill \\   \frac{{\partial \dot m}} {{\partial m}} =  - \left( {\alpha _m  + \beta _m } \right) \hfill \\   \frac{{\partial \dot h}} {{\partial h}} =  - \left( {\alpha _h  + \beta _h } \right) \hfill \\  \end{gathered}  \end{equation}

The terms are given in Section 2.1. The input $u$ and the matrix
$B$ is:

\begin{equation}{\mathbf{B}} = \left[ {\begin{array}{*{20}c}    {\frac{1} {{C_m }}\left( { - g_{Na} m^3 h - g_K n^4  - g_L } \right)}  \\    0  \\    0  \\    0  \\
 \end{array} } \right],\,{\mathbf{u}} = V_E  \end{equation} 

In the next, the repetitive firing condition should be found which
is done through bifurcation analysis using the MATCONT toolbox of
MATLAB \textregistered{}. For this purpose, we will analyze the case
where there is a physiological condition on the sodium channel. That
is the deviation in the equilibrium (rest) potential of the channel
in which MATCONT provides the following data:

\begin{table}
\caption{The repetitive firing condition as a result of deviation in the sodium
channel}

\noindent \centering{}\begin{tabular}{|c|c|}
\hline 
\selectlanguage{english}%
\inputencoding{latin9}Parameter \& Prop & \selectlanguage{english}%
$V_{Na}  = 199.134$ \tabularnewline
\hline
\hline 
\selectlanguage{english}%
\inputencoding{latin9}Eigenvalues & \selectlanguage{english}%
$\begin{gathered}   \lambda _1  =  - 5.04011 \hfill \\   \lambda _2  =  - 0.1257 \hfill \\   \lambda _3  = j0.3958 \hfill \\   \lambda _4  =  - j0.3958 \hfill \\  \end{gathered} $ \tabularnewline
\hline 
\selectlanguage{english}%
\inputencoding{latin9}Equilibrium Points & \selectlanguage{english}%
$\begin{gathered}   V = 0.93404938 \hfill \\   n = 0.33208269 \hfill \\   m = 0.059059 \hfill \\   h = 0.5631245 \hfill \\  \end{gathered} $ \tabularnewline
\hline 
\selectlanguage{english}%
\inputencoding{latin9}Type of Bifurcation & \selectlanguage{english}%
Hopf\tabularnewline
\hline
\end{tabular}
\end{table}

In order to continue the application, a washout filter should be proposed.
The input of the washout filter should be a measurable variable which
is in this case the membrane potential of the nerve fiber. So the
washout filter will be of the following form:

\begin{equation} \begin{gathered}   \dot z = a_w z + b_w V \hfill \\   y = a_w z + b_w V \hfill \\  \end{gathered}  \end{equation} 

In order to have a stable filter the condition \[ a_w  < 0 \]  should
be satisfied. For a practical case one can choose $a_w  =  - 0.01$ 
and $b_w  = 1$ . The resultant filter is augmented into the linearized
nerve model in (2.17) - (2.21) to obtain:

\begin{equation}\begin{gathered}   {\mathbf{\dot x}}_f  = {\mathbf{A}}_f {\mathbf{x}}_f  + {\mathbf{B}}_f V_E  \hfill \\   {\mathbf{A}}_f  = \left. {\left[ {\begin{array}{*{20}c}    {\frac{{\partial \dot V}} {{\partial V}}} & {\frac{{\partial \dot V}} {{\partial n}}} & {\frac{{\partial \dot V}} {{\partial m}}} & {\frac{{\partial \dot V}} {{\partial h}}} & 0  \\    {\frac{{\partial \dot n}} {{\partial V}}} & {\frac{{\partial \dot n}} {{\partial n}}} & 0 & 0 & 0  \\    {\frac{{\partial \dot m}} {{\partial V}}} & 0 & {\frac{{\partial \dot m}} {{\partial m}}} & 0 & 0  \\    {\frac{{\partial \dot h}} {{\partial V}}} & 0 & 0 & {\frac{{\partial \dot h}} {{\partial h}}} & 0  \\    {b_w } & 0 & 0 & 0 & {a_w }  \\
 \end{array} } \right]} \right|_{V_E } ,\,{\mathbf{B}}_f  = \left[ {\begin{array}{*{20}c}    {\mathbf{B}}  \\    0  \\
 \end{array} } \right],\,{\mathbf{x}}_f  = \left[ {\begin{array}{*{20}c}    V  \\    n  \\    m  \\    h  \\    z  \\
 \end{array} } \right] \hfill \\  \end{gathered}  \end{equation}

Numerical representation of the above matrix for the values given
in the Table 2 is now:

\begin{equation}\left[ {\begin{array}{*{20}c}    {\dot V}  \\    {\dot n}  \\    {\dot m}  \\    {\dot h}  \\    {\dot z}  \\
 \end{array} } \right] = \left[ {\begin{array}{*{20}c}    {{\text{ - 0}}{\text{.8261}}} & {{\text{ - 74}}{\text{.9539}}} & {{\text{154}}{\text{.0073}}} & {{\text{5}}{\text{.3840}}} & 0  \\    {{\text{0}}{\text{.0029}}} & {{\text{ - 0}}{\text{.1850}}} & 0 & 0 & 0  \\    {{\text{0}}{\text{.0278}}} & 0 & {{\text{ - 4}}{\text{.0361}}} & 0 & 0  \\    {{\text{ - 0}}{\text{.0042}}} & 0 & 0 & {{\text{ - 0}}{\text{.1186}}} & 0  \\    1 & 0 & 0 & 0 & { - 0.01}  \\
 \end{array} } \right]\left[ {\begin{array}{*{20}c}    V  \\    n  \\    m  \\    h  \\    z  \\
 \end{array} } \right] + \left[ {\begin{array}{*{20}c}    {{\text{ - 0}}{\text{.8261}}}  \\    0  \\    0  \\    0  \\    0  \\
 \end{array} } \right]V_E  \end{equation}

The output feedback will be taken from the output of the washout filter
so the output feedback matrix is $ {\mathbf{C}} = \left[ {\begin{array}{*{20}c}    1 & 0 & 0 & 0 & { - 0.01}  \\
 \end{array} } \right] $. In order to proceed, a full state feedback controller is necessary
for the linear system presented in (2.23). This can be obtained using
the linear quadratic theory (LQR) and the MATLAB\textquoteright{}s
lqr(A, B, Q, R) command directly results a full state feedback coefficient
matrix ${\mathbf{K}}_f $  where $V_E  =  - {\mathbf{K}}_f {\mathbf{x}}$ .
The obtained gain matrix ${\mathbf{K}}_f $  minimizes the infinite
horizon quadratic gain index shown below:

\begin{equation}J = \int\limits_0^\infty  {\left( {{\mathbf{x}}^T {\mathbf{Qx}} + {\mathbf{u}}^T {\mathbf{Ru}}} \right)} dt \end{equation} 

Taking ${\mathbf{Q}} = q{\mathbf{I}}_{4 \times 4} $  and ${\mathbf{R}} = 1$ 
considering the problem of this research the above index is rewritten
as:

\begin{equation}J = \int\limits_0^\infty  {\left( {q\left[ {V^2  + n^2  + m^2  + h^2 } \right] + V_E^2 } \right)} dt \end{equation} 

Substituting $q=100$ and invoking the necessary command the full
state feedback gain coefficient yields:

\begin{equation} {\mathbf{K}}_f  = \left[ {{\text{  - 10}}{\text{.5426   89}}{\text{.5670  - 133}}{\text{.4628    - 6}}{\text{.4330    - 9}}{\text{.8885}}} \right] \end{equation}

The closed loop spectrum $\left( {{\mathbf{A}}_f  - {\mathbf{B}}_f {\mathbf{K}}_f } \right)$ 
for the new case is:

\begin{equation}{\mathbf{\Lambda }}_f  = \left[ {\begin{array}{*{20}c}    {{\text{ - 8}}{\text{.8641}}} & 0 & 0 & 0 & 0  \\    0 & {{\text{ - 3}}{\text{.7002}}} & 0 & 0 & 0  \\    0 & 0 & {{\text{ - 1}}{\text{.0170}}} & 0 & 0  \\    0 & 0 & 0 & {{\text{ - 0}}{\text{.1849}}} & 0  \\    0 & 0 & 0 & 0 & {{\text{ - 0}}{\text{.1186}}}  \\
 \end{array} } \right] \end{equation}

And the corresponding eigenvectors are:

\begin{equation} {\mathbf{V}}_f  = \left[ \begin{gathered}   {\text{ 0}}{\text{.9937    - 0}}{\text{.9621    0}}{\text{.7096    0}}{\text{.0253    0}}{\text{.0011}} \hfill \\   {\text{ - 0}}{\text{.0003    0}}{\text{.0008    - 0}}{\text{.0024    0}}{\text{.9891    0}}{\text{.0000}} \hfill \\   {\text{ - 0}}{\text{.0057    - 0}}{\text{.0796    0}}{\text{.0065    0}}{\text{.0002    0}}{\text{.0000}} \hfill \\   {\text{ 0}}{\text{.0005    - 0}}{\text{.0011    0}}{\text{.0033    0}}{\text{.0016    - 1}}{\text{.0000}} \hfill \\   {\text{ - 0}}{\text{.1122    0}}{\text{.2607    - 0}}{\text{.7046    - 0}}{\text{.1448    - 0}}{\text{.0097}} \hfill \\  \end{gathered}  \right] \end{equation} 

The projective control law will make feedback only from the washout
filter thus the dimension of the feedback is one and the number of
guaranteed retainable eigenvalues is also equal to one. So it is best
to retain the eigenvalue farthest from the imaginary axis which is
the one at -8.8641 so the matrix \[ {\mathbf{V}}_q  \]  is only a
single column vector and found as:

\begin{equation} {\mathbf{V}}_q  = \left[ \begin{gathered}   {\text{ 0}}{\text{.99367}} \hfill \\   {\text{ - 0}}{\text{.00032859}} \hfill \\   {\text{ - 0}}{\text{.0057199}} \hfill \\   {\text{ 0}}{\text{.00048023}} \hfill \\   {\text{ - 0}}{\text{.11223}} \hfill \\  \end{gathered}  \right] \end{equation}

Applying the projection equation in (2.12) gives the feedback gain
from the washout filter output:

\begin{equation}K_o  = {\mathbf{KV}}_q \left( {{\mathbf{CV}}_q } \right)^{ - 1}  = {\text{ - 8}}{\text{.6805}}\end{equation}

And applying the feedback as ${\mathbf{A}}_f  - {\mathbf{B}}_f K_o {\mathbf{C}}$ 
yields the output feedback closed loop eigenvalues:

\begin{equation}\left[ \begin{gathered}   {\text{ - 8}}{\text{.8641}} \hfill \\   {\text{ - 3}}{\text{.1436}} \hfill \\   {\text{ - 0}}{\text{.0013609}} \hfill \\   {\text{ - 0}}{\text{.21673}} \hfill \\   {\text{  - 0}}{\text{.12075}} \hfill \\  \end{gathered}  \right] \end{equation}

So the resultant closed loop is a stable system with the expected
eigenvalue at -8.8641 in the right place. The control law together
with the washout filter is:

\begin{equation} \begin{gathered}   \dot z =  - 0.01z + V \hfill \\   V_E  =  - K_o \left( { - 0.01z + V} \right) \hfill \\  \end{gathered}  \end{equation}

\section{Results and Discussion}

The uncontrolled nerve fiber model in (2.1) with the parameters in
(2.2) except the deviated parameter which is given in Table 2 will
yield a repetitive firing action potential as shown in Figure 1:

\begin{figure}
\selectlanguage{english}%
\noindent \inputencoding{latin9}\includegraphics[scale=0.6]{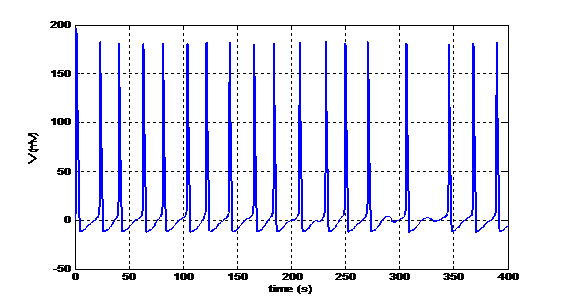}

\inputencoding{latin1}\caption{\selectlanguage{english}%
The repetitive firing response of the bifurcating nerve fiber}

\selectlanguage{english}%

\end{figure}

When the control law of (2.32) is applied through the external voltage
input of the model in (2.3) the response of the nerve fiber becomes
that of the Figure 2. 

\begin{figure}
\selectlanguage{english}%
\noindent \inputencoding{latin9}\includegraphics[scale=0.6]{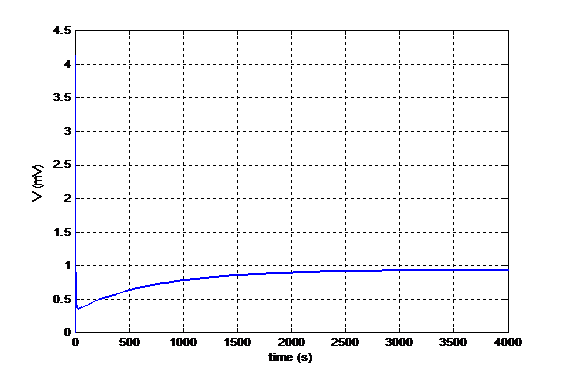}

\inputencoding{latin1}\caption{The stabilized response of the nerve fiber}

\end{figure}

The required level of electrical field can be obtained using the inversion
system presented in (2.7) with the parameters given in Table 1. Required
level of electric field is presented in Figure 3.

\begin{figure}
\selectlanguage{english}%
\noindent \inputencoding{latin9}\includegraphics[scale=0.6]{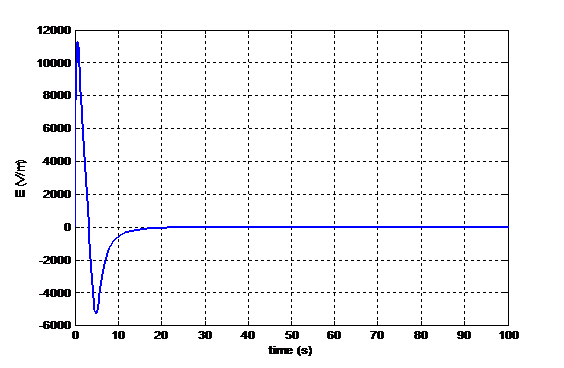}

\inputencoding{latin1}\caption{The required level of electric field for stabilizing the nerve fiber
(only the first 100 seconds are shown where it stays at zero level
after all)}

\selectlanguage{english}%

\end{figure}

As it is expected the level of electric field goes to zero since it
stabilized the equilibrium condition on the nerve fiber. The steady
state portion of the electric field is not allowed to pass from the
washout filter and thus it stays at zero forever after the stabilization.
This makes the electric field stimulation of the nerve fiber very
short. This is a good advantage because of the fact that the level
of electric field required is quite high as seen from Figure 3. In
this research, we have presented a stabilization scheme for stopping
the repetitive firing of the nerve fiber. In order to achieve the
goal, the Hodgkin \textendash{} Huxley model of the squid giant axon
is taken into consideration. A condition on the sodium channel is
found where the fiber starts to produce repetitive firing and membrane
potential is oscillating. After that, the model is linearized and
augmented with a washout filter to obtain a high pass filtered structure.
Finally the projective control approach is utilized to produce a static
output negative feedback on the washout filter output to finish the
application. The simulation results show that the control laws successfully
stabilized the bifurcating nerve fibers with a short lasting electric
field exertion on the fiber. The disadvantage of the method itself
is that it requires the bifurcating conditions to be known. If a parameter
identification method is proposed for the Hodgkin \textendash{} Huxley
model by the measurement of the membrane potential for various types
of nerve fibers. So if the model is properly identified, the research
can be beginning point in the treatment of diseases based on repetitive
firing. In the reverse direction, a stable fiber can be forced to
repetitive firing condition by using approaches like this. For example,
if the dendrite and soma membrane potentials can be separately measured
a pair of pure complex conjugate eigenvalues can be placed so that
the system is forced into a Hopf bifurcation condition. This will
be required since in both Hopf and saddle node bifurcations a pair
of eigenvalue at the specified properties are required.

\end{document}